\documentclass{IEEEtran}
\usepackage{fancyvrb}
\usepackage{url}
\usepackage{tikz}
\begin{document}
 \author{\IEEEauthorblockN{Bernd Ulmann}, 
 \IEEEmembership{Senior Member, IEEE}\\
 \IEEEauthorblockA{FOM University of Applied Sciences\\
  anabrid GmbH\\
  Frankfurt/Main, Germany\\
  Email: ulmann@anabrid.com}\thanks{The author would like to thank his
  colleagues at anabrid GmbH for their incredible work. He would also like
  to thank Mrs. \textsc{Nicole Matje} for proofreading.}}
 \title{Reconfigurable Analog Computers}
 \maketitle
 \begin{abstract}
  The \textsc{Achilles} heel of classic analog computers was the complex, 
  error prone, and time consuming process of programming. This typically 
  involved manually patching hundreds or even thousands of connections between
  individual computing elements as well as setting many precision 10-turn 
  potentiometers manually, often taking hours, or even days. Albeit being 
  simplified by means of removable patch panels, switching from one program to 
  another still was time consuming and thus expensive. 

  With digital computers about to hit physical boundaries with respect to energy
  consumption, clock frequency, and integration density, analog computers have
  gained a lot of interest as co-processors for certain application areas in 
  recent years. This requires some means for automatic reconfiguration of these 
  systems under control of an attached digital computer. 

  The following sections give an overview of classic and modern approaches 
  towards such \emph{autopatch} systems.
 \end{abstract}
 \section{Introduction and requirements}
  Analog computing has been gaining a lot of interest in recent years due to 
  its promise of substantial speedups and unmatched energy efficiency for the 
  solution of problems described 
  by systems of coupled (nonlinear) differential equations or by partial
  differential equations. In contrast to today's prevailing digital computers,
  which solve problems by executing a sequence of rather simple instructions
  read from some memory attached to a central processing unit (\emph{CPU}), 
  an analog computer does not have any memory in the classic sense at all, it
  does not execute instructions, and it typically does not represent values 
  by sequences of bits. Instead, an analog computer works by setting up an 
  (electronic) model of the problem under consideration by connecting a number
  of computing elements in a suitable fashion. These work on values represented
  by continuous voltages or currents and work in continuous time and full 
  parallelism (see \cite{ulmann_ac2} for a thorough treatment of 
  analog and hybrid computers, their history, programming techniques, etc.).

\begin{figure}
   \centering
   \begin{tikzpicture}[
     > = stealth, 
     shorten > = 1pt, 
     auto,
     node distance = 3cm, 
     semithick 
    ]
    \tikzstyle{every state}=[
     draw = black,
     thick,
     fill = white,
     minimum size = 4mm
    ]
    \node (B) at (0, 0) {$b$};
    \node (C) at (0, 1) {$c$};
    \node (PLUS) at (2, 0.5) [circle, draw] {$+$};
    \draw[->] (B) to (PLUS);
    \draw[->] (C) to (PLUS);
    \node (MULT) at (4, .5) [circle, draw] {$*$};
    \draw[->] (PLUS) to (MULT);
    \node (A) at (2, 2) {$a$};
    \draw [->] (A) to (MULT);
    \node (RESULT) at (6, .5) {$x$};
    \draw[->] (MULT) to (RESULT);
   \end{tikzpicture}
   \caption{Computing $a(b+c)$ on an analog computer}
   \label{pic_example}
  \end{figure}
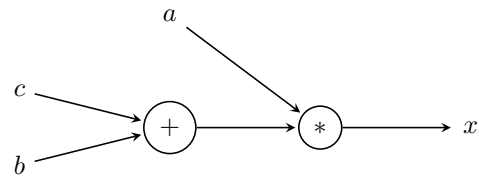

  This different approach is best described by a toy problem such as the 
  computation of $a(b+c)$. On a digital computer this would require six 
  individual instructions: Three to read the values $a$, $b$, and $c$, one 
  addition, one multiplication, and one instruction to store the result back 
  into memory. Since all of these instructions depend on each other there is 
  not too much that can be done in order to achieve parallelism in execution.
  On an analog computer, this problem would require two computing elements,
  one adder and one multiplier connected as shown in figure \ref{pic_example}.
  All values are represented by a corresponding voltage (or a current), so 
  each connection between computing elements is just a single wire, the program
  being a directed graph with edges resembling connections and vertices 
  representing computing elements. 

  It should be noted here that while a digital computer is always capable of 
  trading time against problem size, given that enough memory is available, an
  analog computer must physically grow with the problem size, i.\,e., there must
  be enough individual computing elements for any given problem. Back in the 
  heyday of analog computing, the 1950s--1960s, this was a problem due to the 
  cost and size of analog computers. Using modern CMOS techniques, it is now
  possible to integrate hundreds or thousands of computing elements on a single
  integrated circuit, allowing next generation analog computers to scale to 
  sizes previously unheard of. 

  This problem of scaling, however, is complemented by a great benefit: While 
  the time to solution on a digital computer typically grows with problem size, 
  often much worse than linearly, typically described by the \emph{Big-O
  notation}, the solution time on an analog computer is basically constant, 
  regardless of the number of computing elements required for the 
  implementation of a particular problem. This is a fundamental advantage of
  analog computers over stored-program digital computers.
 
\begin{figure}
   \centering
   \includegraphics[width=\linewidth]{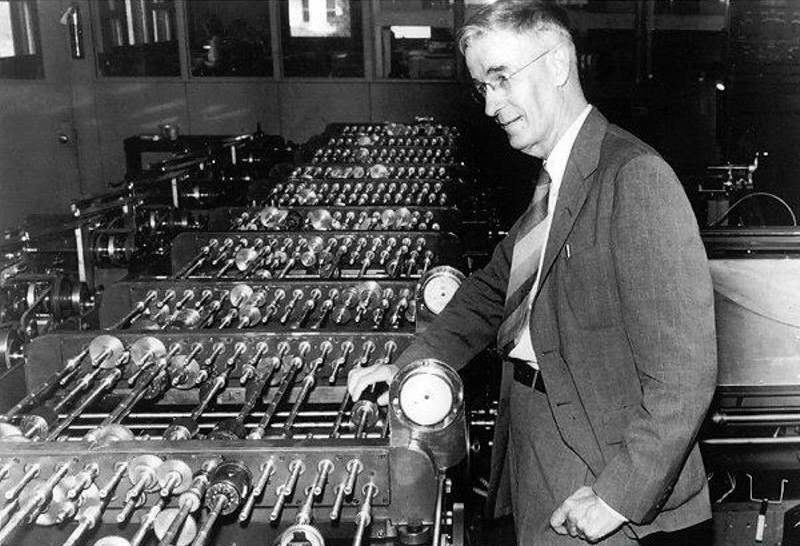}
   \caption{Mechanical differential analyzer}
   \label{pic_bush}
  \end{figure}

\begin{figure}
   \centering
   \includegraphics[width=\linewidth]{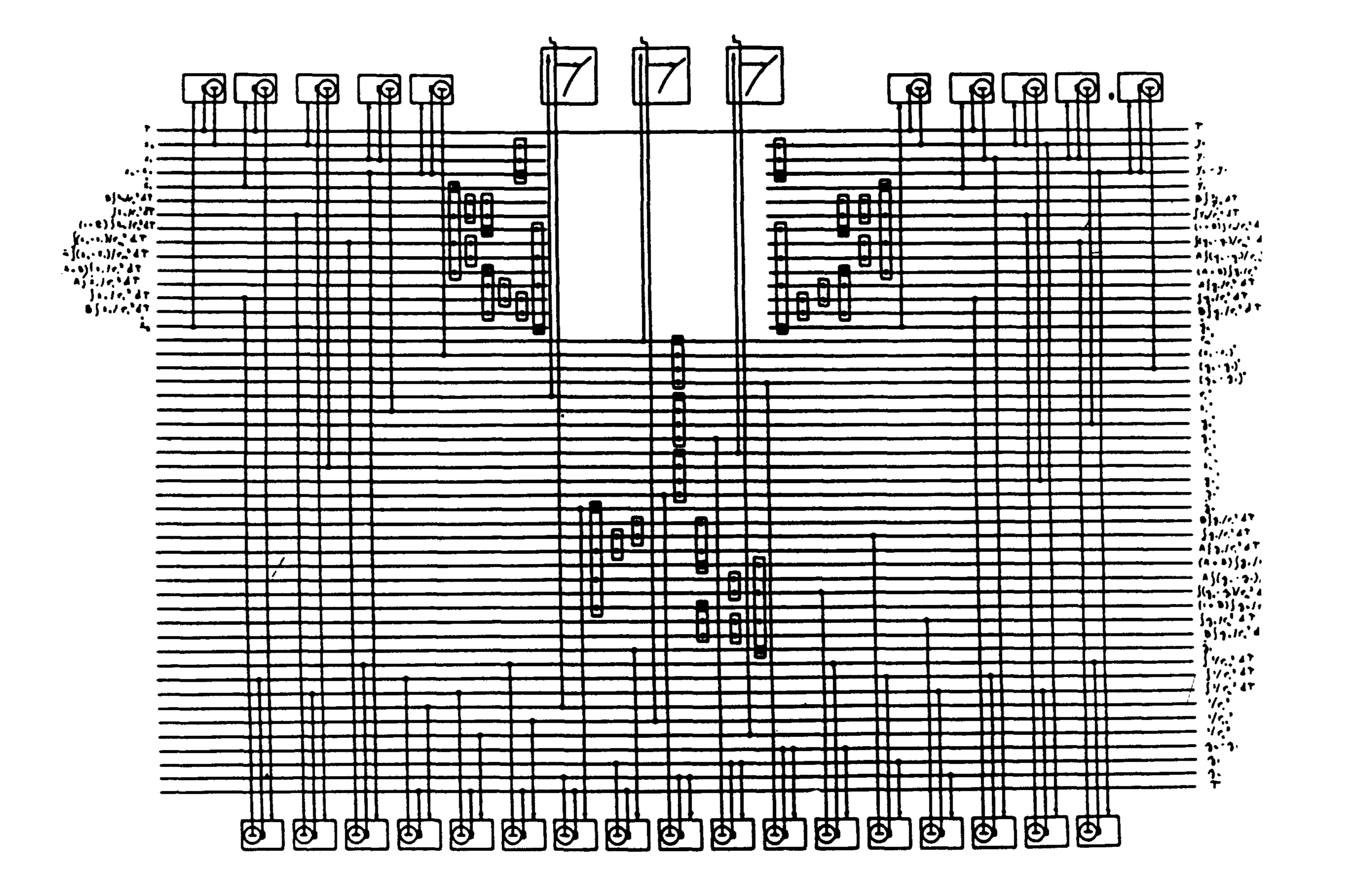}
   \caption{Setup of the \textsc{Bush} differential analyzer for a cosmic
    ray problem (see \cite[p.~77]{owens})}
   \label{pic_vallarta}
  \end{figure}

  Figure \ref{pic_bush} shows an early mechanical differential analyzer
  (a mechanical analog computer) with its creator, \textsc{Vannevar 
  Bush} (11.03.1890--30.06.1974) standing next to it. He is looking 
  at the central interconnect structure. The computing elements are arranged
  to the left and right of this interconnect. This also becomes visible in the
  classic program circuit shown in figure \ref{pic_vallarta}. Variables were 
  transmitted between computing elements by rotating shafts and gears.

  Programming such a machine basically required to dismantle and reassemble it 
  from scratch, a process that could take up to two ``man-days'' (see 
  \cite[p.~277]{bush}). This often invalidated the advantage in speed offered
  by the system as one could also solve the problem executing a manual numeric
  integration technique in about the same time. 
 \section{Early attempts at an autopatch}
  Accordingly, the process of programming the machine, i.\,e., connecting the 
  various computing elements, had to be sped up considerably. This led to the 
  \textsc{Rockefeller} differential analyzer, which still relied on mechanical
  computing elements. However, these were no longer directly connected to 
  each other by means of rotating shafts and gears but used servo motors and 
  synchros for input and output. This made it possible to use a crossbar 
  switch from a telephone exchange system as the central interconnect of the 
  machine, implementing an $n\times m$ matrix of switches.

  From its structure this is the holy grail of analog computing -- having a 
  full $n\times m$ switch matrix makes it possible to connect the computing
  elements without any restrictions. Unfortunately, this does obviously not 
  scale. In a machine such as the \textsc{Rockefeller} analyzer with several
  dozen computing elements it was a viable option, but for machines employing
  hundreds, thousands, of even more computing elements, a full matrix 
  interconnect is just not practical.
 \section{Hannauer's work}
  In 1961, \emph{Electronic Associates Inc.} (\emph{EAI}) performed a study on 
  an autopatch system based on an EAI 231R analog computer. This vacuum tube 
  system featured about $2\cdot 10^2$ individual computing elements. The study
  concluded that a viable switch matrix would require between $5\cdot 10^4$ and 
  $10^5$ individual switches (reed relays back than) -- clearly impractical. 

  In the second half of the 1960s, \textsc{George Hannauer}, also at EAI,
  performed a 
  comprehensive study on the problem of autopatch systems (see \cite{hannauer})
  under a NASA contract. In order to reduce the number of switches required 
  for a realistically large analog computer, a hierarchy of switch matrices
  had to be employed. For this, \textsc{Hannauer} coined the terms 
  \emph{concentrator} and \emph{expander}
  for switch matrices depending on the quotient $e=m/n$ of their inputs and 
  outputs with $e<1$ being a concentrator and $e>1$ an expander.
  \begin{figure}
   \centering
   \includegraphics[width=\linewidth]{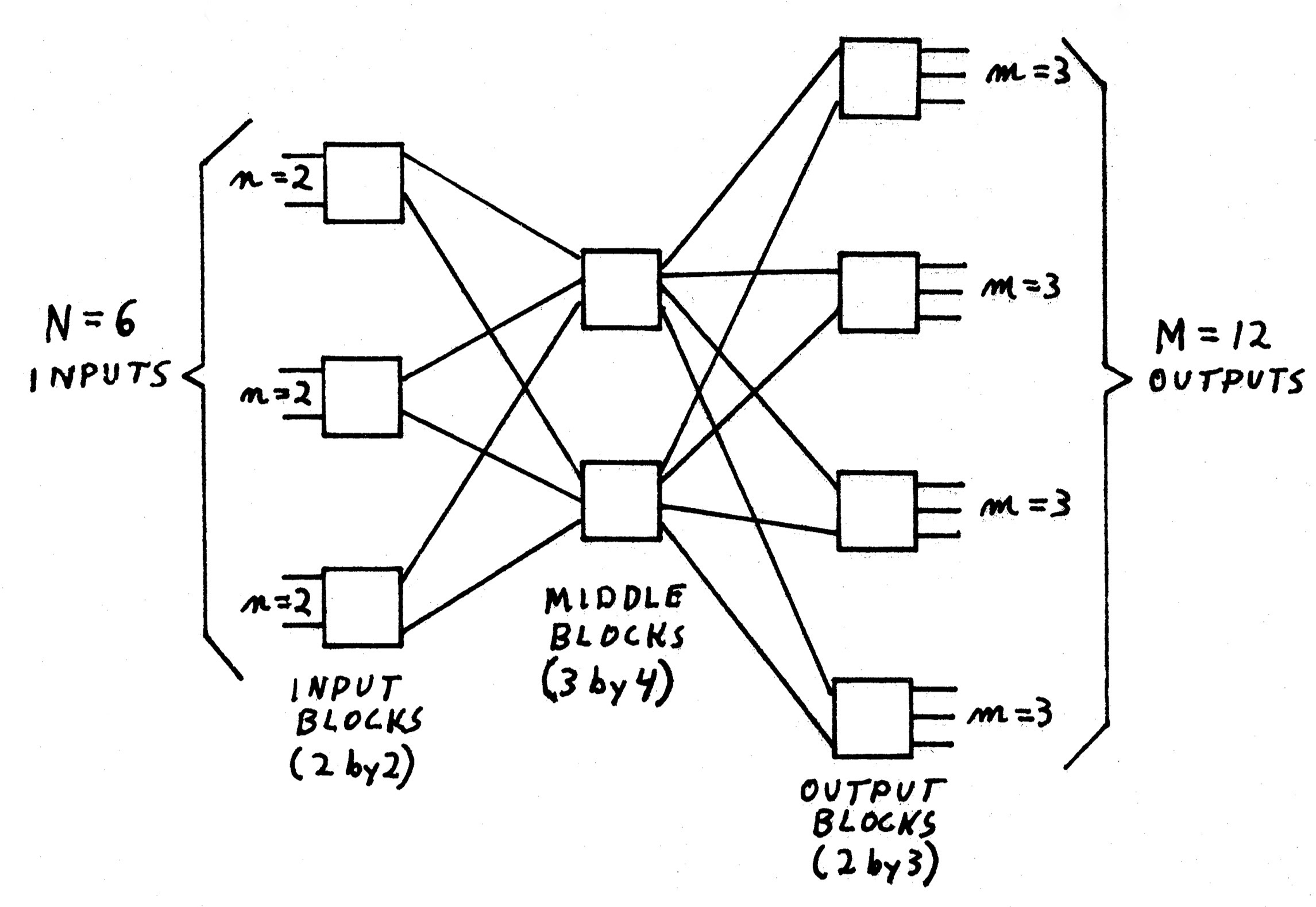}
   \caption{Interconnect structure as proposed in \cite[p.~4-2]{hannauer}}
   \label{pic_hannauer_structure}
  \end{figure}

  This quite closely resembled the structure of a telephone exchange where the 
  number of users substantially exceeds the average number of connections 
  required (see \cite{clos} and \cite{benes} for details). While blocking 
  conditions in a telephone exchange are clearly undesirable, they can be 
  typically resolved by retrying a failed connection attempt. In an analog 
  computer, however, a blocking condition is fatal as it implies that a certain 
  program cannot be implemented at all. 

  Based on this interconnect structure it took EAI about fifteen years to 
  unveil the \emph{SIMSTAR} in 1983. This fully reconfigurable analog computer
  employed the three stage switch matrix structure shown in figure 
  \ref{pic_hannauer_structure}. It consisted of twenty
  $16\times 20$ input blocks for a total of $320$ input signals, twenty 
  $20\times 32$ middle blocks, and $32$ $22\times16$ output blocks yielding
  $512$ output signals. This setup required $30464$ FET switches instead of 
  $163840$ which would be required in the na\"ive approach using a single 
  matrix consisting of $320\times512$ individual switches. SIMSTAR was an 
  incredible technological success but unfortunately a commercial failure as
  it came at a time when digital computers became faster and cheaper at an 
  ever increasing rate.
 \section{Voltage and current coupling}
  The representation of variables by voltages or currents has a substantial
  effect
  on the setup of an analog computer program. Using voltages, as in the vast
  majority of classic and modern analog computers, one output signal of a 
  computing element can be distributed to the inputs of a number of other
  computing elements. However, this requires multiple inputs per computing
  elements -- integrators or summers typically have somewhere between four and
  six inputs, with built-in weights of $1$ and $10$. Accordingly, there are 
  much more computing element inputs than outputs in a voltage coupled system.

  Using currents instead of voltages allows to implicitly sum a number of 
  input currents at the input of a computing element such as an integrator
  or summer, reducing the number of inputs in the system substantially as
  every computing element only has one input (except multipliers or 
  comparators which still require two or more different inputs).
  Unfortunately, an output current cannot be distributed to a number of 
  inputs of following computing elements directly, thus requiring multiple 
  buffered current outputs per computing element.

\begin{figure}
   \centering
   \includegraphics[width=\linewidth]{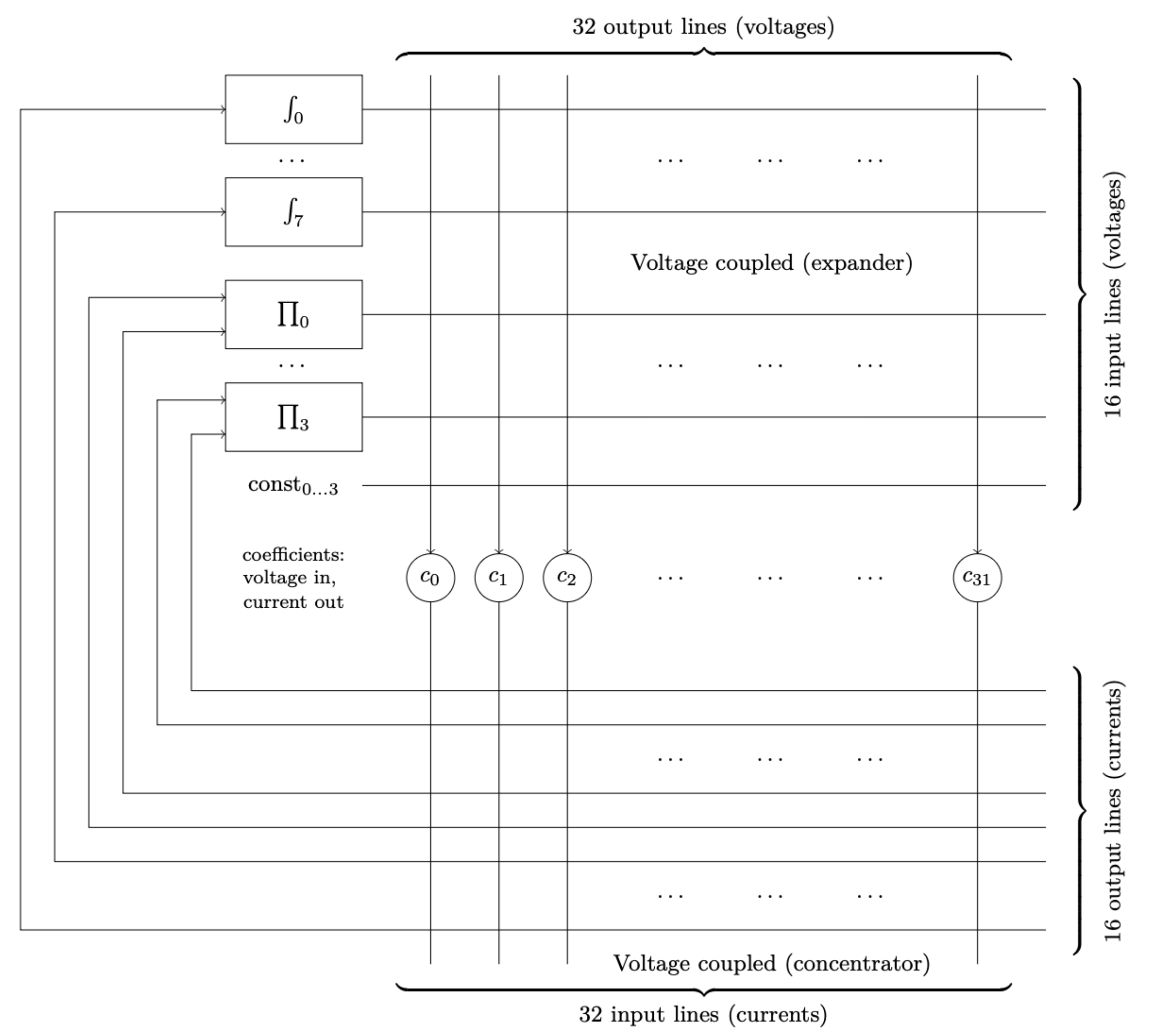}
   \caption{Interconnect system of a LUCIDAC}
   \label{pic_lucidac}
  \end{figure}

  These particular advantages and disadvantages of voltage and current coupling
  can be combined favorably as demonstrated in modern systems such as the 
  small-scale \emph{LUCIDAC},\footnote{See
  \url{https://anabrid.com/lucidac}, retrieved 26.09.2025.} featuring eight
  integrators, four multipliers, and $32$ coefficients or the much larger
  \emph{REDAC},\footnote{See \url{https://anabrid.com/redac}, retrieved 
  26.09.2025.} with up to $1000$ integrators, $500$ multipliers, and $8000$
  coefficients. Figure \ref{pic_lucidac} shows the basic interconnect structure
  of the small LUCIDAC system.

  On the top right is a voltage coupled $16\times32$ switch matrix capable of 
  connecting each input row line to any combination of its $32$ column lines. 
  These column lines each feed a multiplying \emph{digital-analog-converter} 
  (\emph{DAC}) implementing coefficient elements. Each of these coefficients 
  multiplies its input voltage by a $12$ bit value representing a value 
  interval of $[-10,10]$. At its output it delivers a current which is then fed 
  into one of the $32$ column lines of the $32\times 16$ switch matrix shown on
  the bottom right. This matrix is now current coupled. Accordingly, it can 
  perform an implicit summation of input currents onto a single output row 
  line. Each of these $16$ row lines is connected to one input of the computing 
  elements. 

  Using this mixed voltage/current value representation, the analog computer 
  does not need any explicit summers as computing elements, thus greatly 
  reducing the number of connections within a given setup. The degree of 
  sparsity in a setup like this is determined by the number of column lines of 
  the two matrices, which can be easily extended beyond the $32$ lanes shown 
  here.

\begin{figure}
   \centering
   \includegraphics[width=\linewidth]{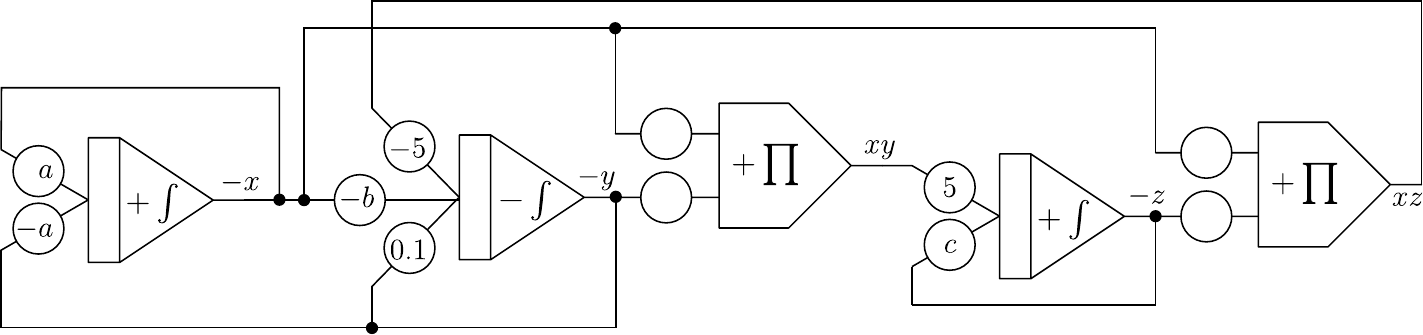}
   \caption{Resulting analog computer setup for the chaotic \textsc{Lorenz} '63
    system}
   \label{pic_lorenz_circuit}
  \end{figure}

  Programming these systems is done using a \emph{domain specific language}
  (\emph{DSL}) and a compiler capable of translating problem descriptions 
  into suitable configuration bitstreams. Figure \ref{pic_lorenz_code} shows
  the input to compiler specifying a chaotic \emph{Lorenz} '63 system (see
  \cite{lorenz}). The resulting analog computer setup is shown in figure
  \ref{pic_lorenz_circuit}. Integrators are represented by triangular 
  structures with a small rectangular box on the left, the elements inscribed 
  with $+\Pi$ are multipliers, and coefficients are represented by circles.
  \begin{figure}
   \centering
   {\scriptsize
   \begin{Verbatim}
fn X(t);
fn Y(t);
fn Z(t);

let diff[X, t] = 1.8 * Y - X;
let diff[Y, t] = 1.56 * X * (1 - 2.678 * Z) - 0.1 * Y;
let diff[Z, t] = 1.5 * X * Y - 0.2667 * Z;

let X(t: 0) = 0.1;
let Y(t: 0) = 0.0;
let Z(t: 0) = 0.0;

plot(x: X(t), y: Y(t));

out X(t);
out Y(t);
   \end{Verbatim}
   }
   \caption{DLS description of a chaotic \textsc{Lorenz} '63 system}
   \label{pic_lorenz_code}
  \end{figure}

  Since there is one coefficient per column line in this particular autopatch
  system, every input to a computing element has an associated coefficient 
  element. Since the inputs to the computing elements are current coupled, 
  multiple input values can be combined at the current input of an element.
  Noteworthy is the absence of any summers from this circuit as summation 
  takes place in the current coupled matrix.
 \section{Coefficients}
  Another interesting aspect of modern reconfigurable analog computers is the 
  implementation of its coefficient elements, namely their resolution and 
  available value interval. While coefficients in classic analog computers,
  being just voltage dividers, were 
  restricted to values within $[0,1]$, using multiplying DACs with output 
  buffers, it is now easy to extend the interval into the negative
  domain such as $[-1,1]$. This greatly simplifies programming and system
  implementation as no inverters
  are needed for sign inversions, further reducing the number of computing
  elements required for a particular problem. 

  Using current coupling, at least in the output stage of a coefficient, makes
  it possible to extend the value interval further with $[-10,10]$ being 
  desirable. This further aids programming and simplifies the overall analog
  computer implementation as no built-in weights for integrators, etc., are
  required to upscale values. 

  In a system such as the LUCIDAC, every column lane connecting the two 
  switch matrices has an associated coefficient. Accordingly, the number of 
  coefficient elements grows with the degree of connectivity, causing 
  substantial cost of implementation due to the complexity of high resolution
  coefficients. 

  However, a study of real analog computer programs shows that not all 
  coefficients need $12$ or more bits of resolution. In fact, a substantial 
  number of coefficients will have values such as $\pm10$, $\pm1$, 
  $\pm0.5$, and $\pm0.1$, requiring only three bits of configuration data per 
  coefficient. The actual proportion of high resolution coefficients to these
  low resolution elements highly depends on the class of problems the analog
  computer is used for, but having about $20$\% to $30$\% of low resolution 
  coefficients seems to be sufficient to implement most if not all problems.
 \section{Configuration time}
  With solution times in the order of several $\mu$s, these modern analog 
  computers pose the same problem as \textsc{Vannevar Bush}'s differential 
  analyzers nearly $100$ years ago: The configuration time must be on par 
  with the solution time. Ideally, it should be even smaller in order to not 
  lose the speed advantage of the actual solution process due to a slow setup. 

  Since the number of configuration bits required for a system scales linearly
  with the number of switches to be controlled and the number of coefficients
  times their respective solution, it will be necessary to implement some 
  way of performing ``sparse configurations'', i.\,e., update switch matrices 
  and coefficient values on a ``need to change'' basis instead of uploading a 
  full set of configuration data every time a setup is changed. 
 \section{Conclusion}
  Analog computers will play a major role in future computer systems as 
  specialized co-processors, alleviating the digital processor from a wide
  range of compute intensive tasks. Elaborate autopatch systems as described
  above are of central importance for this particular applications. 

  In addition to this, suitable software support in order to abstract
  from the underlying analog computing paradigm as much as possible and to 
  integrate this class of co-processors seamlessly into existing software
  systems is required, too. This will include libraries for popular programming
  languages such as Python as well as plug-ins for systems like MATLAB, etc.

  These developments will bring analog computers back to a wide variety of 
  applications, most noteworthy \emph{high-performance-computing} (\emph{HPC})
  where their extremely high energy efficiency and fast solution times will 
  make it possible to tackle problems currently not within reach of classic
  digital computers.
 \begin{IEEEbiography}[{\includegraphics[width=1in,height=1.25in,clip,keepaspectratio]{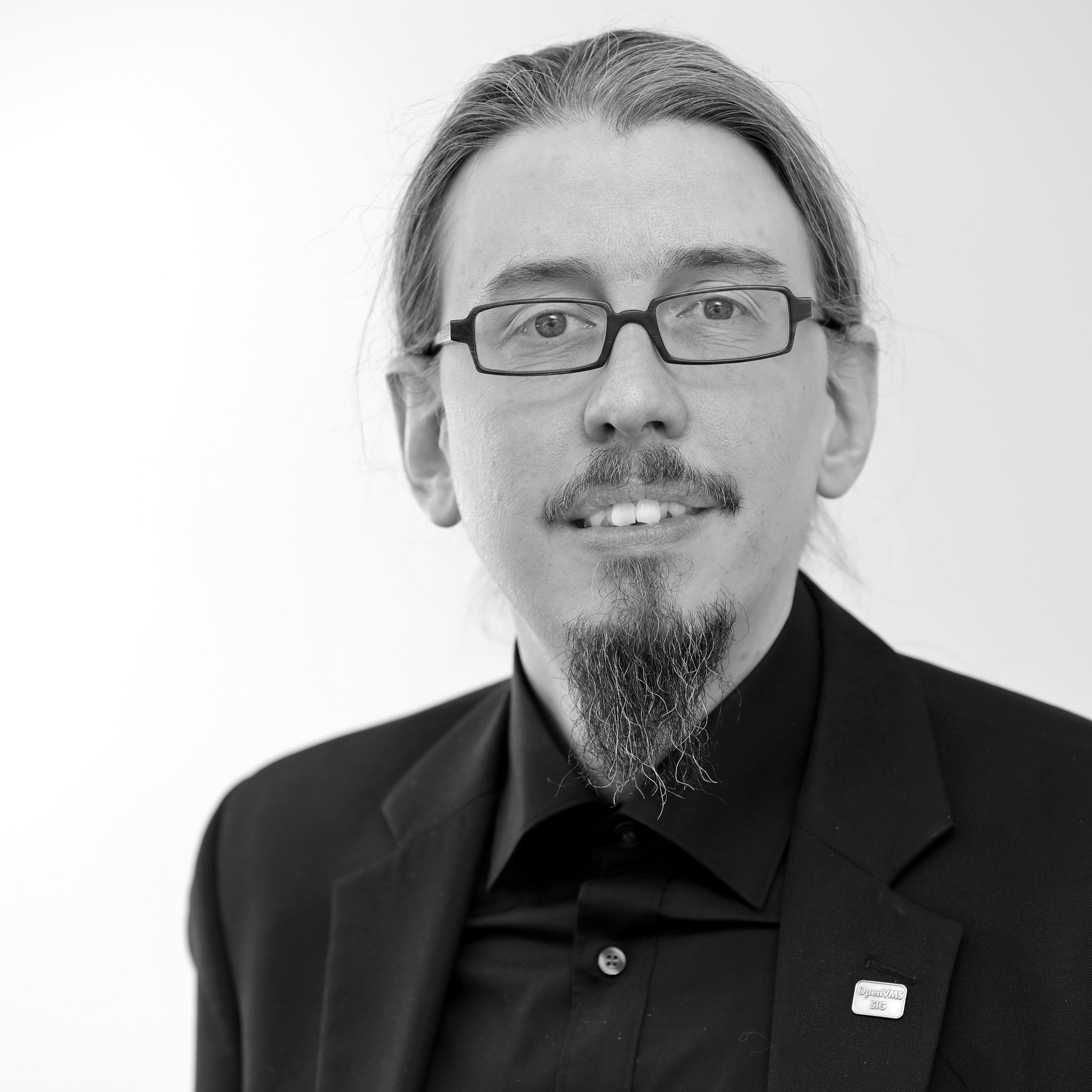}}]{Bernd Ulmann} was born in Neu-Ulm, Germany in
  1970. He received his diploma in mathematics from the Johannes 
  Gutenberg-Universit\"at, Mainz, Germany, in 1996. He received his Ph.D. from
  the Universit\"at Hamburg, Germany, in 2009. 
  
  He is professor for business informatics at the FOM University of
  Applied Sciences, Frankfurt/Main, Germany. His main interests are analog and 
  hybrid computing, the simulation of dynamic systems and analog 
  implementations of chaotic systems. He is author of several books on these
  topics.
 \end{IEEEbiography}
 \vfill
\end{document}